\documentclass[floats,preprint,prl,aps]{revtex4}
\usepackage{graphicx}

\newcommand\beq{\begin{equation}}
\newcommand\eeq{\end{equation}}
\newcommand\bea{\begin{eqnarray}}
\newcommand\eea{\end{eqnarray}}
\newcommand\non{\nonumber}
\newcommand\bib{\bibitem}

\begin{document}


\title{\bf Quantum field theoretical study of an effective spin model in coupled  
optical cavity arrays}

\author{\bf Sujit Sarkar}
\address{\it Poornaprajna Institute of Scientific Research,
4 Sadashivanagar, Bangalore 5600 80, India.\\
E-Mail: sujit@physics.iisc.ernet.in\\
Phone: 091 80 23612511/23619034, Fax: 091-80-2360-0228 \\}
\date{\today}

\begin{abstract}
Atoms trapped in microcavities and interacting through the exchange of virtual
photons can model an anisotropic Heisenberg spin-1/2 lattice.
 We do the quantum field theoretical study of such a system using the
Abelian bosonization method followed by the renormalization group analysis.
We present interesting physics due to the presence of exchange anisotropy.
An infinite order Kosterliz-Thouless-Berezinskii transition is replaced
by second order XY transition even an infinitesimal  
a small anisotropy in exchange coupling is introduced.
We predict a quantum phase transition between Mott insulating and photonic 
superfluid
phase due to detuning between the cavity and laser frequency,
a large detuning favours the 
photonic superfluid phase. We also do the analysis of Jaynes and Cumming
Hamiltonian to support results of quantum field theoretical study.

Pacs: 42.50.Dv, 42.50.Pq, 03.67.Bg, 75.10.Jm  
  
\end{abstract}
\maketitle


\section{I. Introduction}
The physics of strongly correlated system is interesting in its own right
and manifests in different branches of physics. Some of the important
correlated physics appears in natural oxide materials \cite{rao} 
and some of them are
in engineered materials, like the correlated physics in 
Josephson junction array \cite{lik1},
Bose-Einstein condensation and optical lattice \cite{dens,zoller}. 
Therefore one can raise
the question what is the further source of correlated physics in the
state of enginnering ?. 
The recent experimental success in engineering strong interaction 
between the photons and atoms in high quality micro-cavities opens up
the possibility to use light matter system as quantum simulators for
many body physics 
\cite{green,hart1,hart2,ji,byrn,caru,bhas,toma,zhao,pipp,rossi,horn,blat}.
\\ 
Here we would like to discuss the basic physics of micro-optical cavity
very briefly.
A micro-cavity can be created in 
a photonic band gap material by producing a localised defect in the structure
of the crystal, in such a way that light of a particular frequency can not
propage outside the defect area. Large arrays of such micro-cavities have been
produce. Photon hopping between neighboring cavities has been observed in the
microwave and optical domains. 
Many body Hamiltonians can be created and probed in coupled cavity arrays. 
There atoms are used for detection and also for the generation of
interaction between photons in the same cavity.
As the distance between the adjacent cavities is considerably
larger than the optical wave length of the resonant mode, individual cavities
can be addressed. These artificial system could act as a quantum simulator.
In this optical cavities system we study the
different quantum phase of polariton ( a combined excitations of atom-photon
interactions. ) by using spin models
that conserve the total number of excitations. Best of our knowledge, at first
we have done the explcit quantum field theoretical calculations of this 
type of system.\\ 

At first we would like to 
discuss the generation of the spin model for such type of systems. It has 
already been discussed in the literature but we mention this process 
very briefly for the sake
of completeness. 
In this description we will follow the Ref. \cite{hart1} 
and Ref. \cite{hart2}. 

Micro-cavities of the photonic crystal are coupled through the exchange of photons.
Each cavity consists of one atom with three levels in the energy
spectrum, two of them are long lived
and represents two spin states of the system and the other is excited. Externally
applied laser and cavity modes couple to each atom of the cavity. It may 
induce the Raman transition between these two long lived levels. Under a
suitable detuning between the laser and the cavity modes virtual photons
are created  
in the cavity which mediate an interaction with another atom in a 
neighboring cavities. One can eliminated the excited states of atomic level
photon states by choosing the appropriate detuning between the applid laser
and cavity modes. 
Then one can achieve only two states per atom in the long 
lived state and the system can be described by a spin-1/2 
Hamiltonian \cite{hart1,hart2}. 

Fig. 1 shows the
schematic phase diagram of our description to generate the xy spin interaction
of the system. The Hamiltonian of the system consists of three parts:
\beq
H ~= ~ {H_A} ~+~{H_B} ~+~ {H_C}
\eeq     
Hamiltonians are following
\beq
{H_A} ~=~ \sum_{j=1}^{N} { {\omega}_e } |e_j > <e_j | ~+~ 
{\omega}_{ab} |b_j > <b_j | 
\eeq
$j$ is the cavity index. ${\omega}_{ab} $ and ${\omega}_{e} $ is 
the energy of the state $ | b> $ and the excited state respectively. The
energy level of state $ |a > $ is set zero. 
The following Hamiltonian  describes photons in the cavity,
\beq
 {H_C} ~=~ {{\omega}_C} \sum_{j=1}^{N} {{a_j}}^{\dagger} {a_j} ~+~
{J_C} \sum_{j=1}^{N} ({{a_j}}^{\dagger} {a_{j+1}} + h.c ),  
\eeq 
where ${a_j}^{\dagger}$ creates a photon in cavity $j $, ${\omega}_C $
is the energy of photons and $ J_C $ is the tunneling rate of photons
between neighboring cavities.
Interaction between the atoms and photons and also by the driving lasers
are described
\beq
{H_{AC}}~=~ \sum_{j=1}^{N} [ (\frac{{\Omega}_a}{2} e^{-i {{\omega}_a} t} +
{g_a} {a_j}) |e_j > < a_j | + h.c] + [a \leftrightarrow b ] . 
\eeq
Here ${g_a} $ and ${g_b} $ are the couplings of the cavity mode for the
transition from energy states $ |a > $ and $ | b> $ to the excited state.
${\Omega}_a $ and ${\Omega}_b $ are the Rabi frequencies of lasers
with frequencies ${\omega}_a $ and $ {\omega}_b $ respectively.

They
have derived an effective spin model by considering the following physical
processes:
A virtual process regarding emission and absorption of
photons between the two stable  states of neghiboring cavities. The resulting 
effective Hamiltonian is
\beq
{H_{xy}} = \sum_{j=1}^{N}  B {{\sigma}_j}^{z} ~+~\sum_{j=1}^{N} 
({J_1} {{\sigma}_j}^{\dagger} {{\sigma}_{j+1}}^{-} ~+~
{J_2} {{\sigma}_j}^{-} {{\sigma}_{j+1}}^{-} + h.c )
\eeq 
When $J_2 $ is real then this Hamiltonian reduces to the XY model.
Where ${{\sigma}_j}^{z} = |b_j > <b_j | ~-~ |a_j > <a_j | $,
${{\sigma}_j}^{+} = |b_j > <a_j | $, ${{\sigma}_j}^{-} = |a_j > <b_j | $
\beq
H_{xy}~=~ \sum_{i=1}^{N} B ( {{\sigma}_i}^{z}~+~{J_x} {{\sigma}_i}^{x}
{{\sigma}_{i+1}}^{x} ~+~ {J_y} {{\sigma}_i}^{y}
{{\sigma}_{i+1}}^{y}) .
\eeq
With ${J_x} = (J_1 + J_2 )/2 $ and ${J_y} = (J_1 - J_2 )/2 $.\\

Here we discuss very briefly about an effective 
$ {{\sigma}_i}^{z} {{\sigma}_i}^{z}$ in such a system. Authors of 
Ref.\cite{hart1,hart2}
have proposed the same atomic level configuration but having only one
laser of frequency ${\omega}$ that mediates atom-atom coupling through
virtual photons. Another laser field with frequency $\nu $ is used to
tune the effective magnetic field. They described the one-dimensional case.
In this case the Hamiltonian ${H_{AC}} $ will change but the Hamiltonians $H_A $
and $H_C $ will not. 
\beq
{H_{AC}}~=~ \sum_{j=1}^{N} [ (\frac{{\Omega}_a}{2} e^{-i {{\omega}_a}t} +
\frac{{\Lambda}_a}{2} e^{-i {{\nu}_a}t}
{g_a} {a_j}) |e_j > < a_j | + h.c] + [a \leftrightarrow b ] .
\eeq
Here, ${\Omega}_a $ and ${\Omega}_b $ are the Rabi frequencies of the
driving laser with frequency ${\omega}$  on transition $|a > \rightarrow |e> $ 
$|b > \rightarrow |e> $, whereas ${\Lambda}_a $ and ${\Lambda}_b $
driving laser with frequency ${\nu}$  on transition $|a > \rightarrow |e> $
$|b > \rightarrow |e> $. They have eliminated adiabatically the excited atomic 
levels and
photons by considering the interaction picture with respect to 
$ H_0 = H_A ~+~H_C $. They have considered the detuning parameter in such
a way that the Raman transitions between two level supressed. They have also
chosen the parameter in such a way that the dominant two-photon processes are
those that involve one laser photon and one cavity photon each but the atom
does no transition between levels a and b. Whenever two atoms exchange a 
virtual photon then both of them experience a Stark shift plays the
role of an efective $ {{\sigma}^{z}}{{\sigma}^{z}} $ interaction. Then
the effective Hamiltonian reduce to 
\beq
{H_{zz}}~=~\sum_{j=1}^{N} ( \tilde{B} {{\sigma}_j}^{z} ~+~ {J_z}
{{\sigma}_j}^{z}{{\sigma}_{j+1}}^{z} )
\eeq
Analytical expressions for $\tilde{B}$, $J_1 $, $J_2$  and $ J_z $ has given 
in Ref. \cite{hart2}.
These two parameters can be tuned independently by varying the laser frequencies.
They have obtained an effective model by combining Hamiltonians $H_{xy} $ and
$H_{zz} $ by using Suzuki-Trotter formalism. The effective Hamiltonian
simulated by this procedure is
\beq
H_{spin} ~=~\sum_{j=1}^{N} ( B_{tot} {{\sigma}_j}^{z} ~+~ 
\sum_{{\alpha}=x,y,z} J_{\alpha} {{\sigma}_j}^{\alpha} {{\sigma}_{j+1}}^{\alpha})
\eeq 
where $ B_{tot} = B + \tilde{B} $.  It has been shown in Ref. \cite{hart2} 
that $J_y $ is less than
than $J_x $. It is clear from analytical expressions for
$J_x $ and $ J_y $ that the magnitudes of ${J_1}$ and $J_2 $ are different. 
This result of numerical simulations trigger us to define a model, 
which has
given below to study the quantum phases of this system and also the
transition among them and at the same
time this subject is in the state of art of engineering. 

\section{II. Renormalization Group study of model Hamiltonian} 
We consider the anisotropic Heisenberg spin-1/2 Hamiltonian on
a one dimensional lattice.
The XYZ Heisenberg Hamiltonians is defined as:
\bea
H_{XYZ} ~=~ \sum_n ~[ & & (1+a) ~S_n^x S_{n+1}^x ~+~ (1-a)~ S_n^y S_{n+1}^y \non \\
& & +~ \Delta ~S_n^z S_{n+1}^z ~+~ h ~S_n^z ~]~,
\label{ham2}
\eea
where $S_n^{\alpha}$ are the spin-1/2 operators.
We assume that the $XY$ anisotropy $a$ and the $zz$ coupling $\Delta$
satisfy $-1 \le \Delta \le 1$, and $ 0 < a \leq 1 $
and magnetic field strength $h \ge 0$.
The Hamiltonian $H_{XYZ}$ is invariant under the
transformation $S_n^x \rightarrow - S_n^x$, $S_n^y \rightarrow - S_n^y$, $S_n^z
\rightarrow -  S_n^z$, actually it is a $Z_2$ symmetry.
For finite $h$,
$Z_2$ symmetry is absent when $S_n^z \rightarrow - S_n^z$.
Here $h \sim B_{tot}$, ${\Delta = J_z}$, ${J_1 =1 }$ and ${J_2} =a $.
\\
Spin operators can be recasted in terms of spinless fermions through
Jordan-Wigner
transformation and then finally one can express the spinless fermions
in terms of bosonic fields \cite{gia}.
We recast the spinless
fermions operators in terms of field operators by this relation \cite{gia}.
\beq
{\psi}(x)~=~~[e^{i k_F x} ~ {\psi}_{R}(x)~+~e^{-i k_F x} ~ {\psi}_{L}(x)]
\eeq
where ${\psi}_{R} (x)$ and ${\psi}_{L}(x) $ describe the second-quantized
fields of right- and
the left-moving fermions respectively.
$k_F$ is Fermi wave vector.
Therefore, one can study the effect of gate voltage
through arbitrary $k_F$.
We would like to express the fermionic fields in terms of bosonic
field by the relation
\beq
{{\psi}_{r}} (x)~=~~\frac{U_r}{\sqrt{2 \pi \alpha}}~
~e^{-i ~(r \phi (x)~-~ \theta (x))},
\eeq
$r$ is denoting the chirality of the fermionic fields,
right (1) or left movers (-1).
The operators $U_r$ commutes with the bosonic field. $U_r$ of different species
commute and $U_r$ of the same species anti-commute.
$\phi$ field corresponds to the
quantum fluctuations (bosonic) of spin and $\theta$ is the dual field of $\phi$.
They are
related by the relations
$ {\phi}_{R}~=~~ \theta ~-~ \phi$ and  $ {\phi}_{L}~=~~ \theta ~+~ \phi$.

Hamiltonian $H_0$ is non-interacting part of $H_{XYZ}$.
\beq
H_0 ~=~ \frac{v}{2} ~\int ~dx ~[~ (\partial_x \theta)^2 ~+~ (\partial_x
\phi)^2 ~]~,
\label{ham3}
\eeq
where $v$ is the velocity of the low-energy excitations. It is one of the Luttinger
liquid parameters and the other is $K$, which is related to $\Delta$
by \cite{gia}
\beq
K ~=~ \frac{\pi}{\pi + 2 \sin^{-1} (\Delta)} ~.
\label{kd}
\eeq
$K$ takes the values 1 and 1/2 for $\Delta =0$ (free field), and $\Delta =1$
(isotropic anti-ferromagnet), respectively.
The relation between $K$ and $\Delta$ is not preserved under the
renormalization, so this relation is only correct for initial
Hamiltonian.
The analytical form of the spin operators
in terms of the bosonic fields are
\bea
S_n^x ~&=&~ [~ c_2 \cos (2 {\sqrt {\pi K}} \phi) ~+~ (-1)^n c_3 ~]~
\cos ({\sqrt {\frac{\pi}{K}}} \theta ), \non \\
S_n^y ~&=&~ -[~ c_2 \cos (2 {\sqrt {\pi K}} \phi) ~+~ (-1)^n c_3 ~]~
\sin ({\sqrt {\frac{\pi}{K}}} \theta ), \non \\
S_n^z ~&=&~ {\sqrt {\frac{\pi}{K}}} ~\partial_x \phi ~+~ (-1)^n c_1
\cos (2 {\sqrt {\pi K}} \phi ) ~,
\label{spin}
\eea
where $c_i$s are constants as given in Ref. \cite{zamo}.
The Hamiltonian $H_{XYZ}$ in terms of bosonic fields is the
following,
\bea
H_{XYZ} &=& H_0 + a \int \cos (2 {\sqrt {\frac{\pi}{K}}}
\theta (x) ) dx + \Delta \int \cos(4 {\sqrt{\pi K}} \phi (x)) dx \non \\
 & & -h \int {{\partial}_x } {\phi (x) } dx 
\label{ops1}
\eea
One can get the $H_{XY}$ Hamiltonian by simply putting $\Delta =0$
in the above Hamiltonian.
In this derivation, different powers of
coefficients $c_i$ have been absorbed
in the definition of $a, h$ and $\Delta$.
The integration of the oscillatory terms in the Hamiltonian yield
negligible small ontributions, the origin of the oscillatory terms are the 
spin operators.
So it's a resonably good approximation to keep only the non-oscillatory
terms in the Hamiltonian
. The Gaussian scaling dimension of these
coupling terms, $a, \Delta$ are $1/K$, $4K$ respectively.
The third term ($ \Delta $) of the Hamiltonian tends to order 
the system into density wave phase, whereas the second term 
($ a $) of the Hamiltonian favours the staggered order in 
the xy plane. Two sine-Gordon couplings terms are from two dual
fields. Therefore the model Hamiltonian consists of two competing
interactions between the ordered phase and the $XY $ order. 
This Hamiltonian contains two 
strongly relevant and mutually nonlocal perturbation over
the Gaussian (critical) theory.
In such situation strong coupling fixed point is usually
determined by the most relevant perturbation whose amplitude
grows up according to its Gaussian scaling dimensions and
it is not much affected by the less relevant coupling terms.
However this is not the general rule, if the two operators
exclude each other, i.e., if the field configurations which
minimize one perturbation term do not minimize the other.
In this case interplay between the two competing relevant
operators can produce a novel quantum phase transition through
a critical point or a critical line. Therefore we would like to
study the RG euation to interpret the quantum phases of the system.\\
We will now study how the parameters $a$, $\Delta$ and $K$ flow under RG.
The operators in Eq. 17 are related to each other
through the
operator product expansion; the RG equations for their coefficients
will therefore be coupled to each other.
Here we derive the RG equations by using perturbative renormalization
group approach scheme. We use operator product expansion to derive
these RG equations which is independent of boundary condition
\cite{cardy}. RG equations themselves have been established
in a perturbative expansion in coupling constant ($g(l)$), they
cease to be valid beyond the certain length scale, where
$g(l) \sim 1$ \cite{gia}.
The RG equations for the coefficients of Hamiltonian $H_{XYZ}$ are
\bea
\frac{da}{dl} ~&=&~ (2 -\frac{1}{K}) a ,\non \\
\frac{d{\Delta}}{dl} ~&=&~ (2 - 4K) {\Delta}  \non \\
\frac{dK}{dl} ~&=&~ \frac{a^2}{4} ~-~ K^2 {O}^2 ~,
\label{rg2}
\eea
These RG equations have trivial (${a^*}= 0 = {{\Delta}^*}$)
fixed points for any arbitrary $K$. 
Apart from that these RG equations have also two non-trivial
fixed lines, $a = \Delta $ and $a = - {{\Delta} }$
for $ K =1/2 $.
In our study, there are critical surfaces on which the system flows onto
the non-trivial fixed lines (${a} = \pm {{\Delta} }$). A density
wave states can be characterized when $K \rightarrow 0 $ or the staggered ordered
when ($ K \rightarrow \infty $). Note that the transition occuring on them
are second order. Infinitesimal amount of anisotropy will change the situation
drastically, a gapless phase in absence of anisotropy will change to 
the gapped phase
in presence of anisotropy. Since this gapped excitation is not directly related
to magnetization, therefore it will not favour to create the plateau phase. When
the system is in the plateau phase the transition driven by the magnetic field
is always of $ z=2 $ ( z is the dynamical critical exponent) 
and thus the plateau shows a square-root behaviour of 
magnetization.
When $a$ is increasing, a second order transition drives system to 
the in plane XY ordered phase
, whose exponents depend on initial couplings and hence are nonuniversal.
In absence of planar anisotropy the transition to plateau state is Kosterliz-
Thouless-Brezinskii (KTB). When the inplane anisotropy is present, 
then $z=1 $ 
\cite{phase}. Please see Refs. \cite{subir,sondhi} for
detail understanding of this subject.
A magnetic field larger than the relevant gap of the system drives
the system to a gapless phase. This transition is from commensurate
phase to incommensurate phase transition.

We have seen the analytical expression of $B_{tot}$ from Ref. \cite{hart2} that
the total magnetic field increase for the larger
values of detuning, therefore  
larger detuning drives the system from gapped (Mott-insulating) state
to gapless superfluid state.\\

 We now discuss, how the effective repulsion
will decrease as we increase the detuning between the atomic and laser 
frequency. It can be explained starting from the 
Jaynes-Cummings Hamiltonian \cite{jay,horo}.
Janes-Cummings Hamiltonian for a single atom is
\beq
H_{JC} ~=~ { {\omega}_C }{ {a}^{\dagger} a} + {{\omega}_0} |e> <e| +
{\lambda} ( {a}^{\dagger} |g><e| ~+~a |e> <g| )
\eeq
${{\omega}_C }$ and ${{\omega}_0}$ are the frequency of the resonant mode
of the cavity and of the atomic transition, respectively. ${\lambda}$ is
the Jaynes-Cumming coupling between the cavity mode and the two level system.
${a}^{\dagger}$ (${a}$) is the creation (annihilation) operator of a photon
inside a cavity. $ |g > $ and $ |e> $ is respectively the ground state and
excited states of the two level system respectively. When we consider large
values of photon and atomic frequencies compare to atom-photon coupling $\lambda$,
the number of excitations is conserved for this Hamiltonian. Suppose
we consider fixed numbers of excitation, $ n$. The energy eigenvalues for 
$n$ excitations is
\beq
{E_n}^{\pm} = n {{\omega}_C} + \frac{\Delta}{2} 
\pm \sqrt{ n {\lambda}^{2}~ +~ \frac{{\Delta}^2 }{4} } 
\eeq
Here $\Delta ~=~ ({\omega}_0 - {\omega}_c )$ and $n \geq 1 $.
Now we consider an array of cavities, the basic Hamiltonian for each cavity
is the same as that of Eq.19 . Here we consider the system with one excitation
of energy $ E_1 $ in each cavity and the lowest energy of two excitation in
each cavity is $ E_2 $. Therefore to create one additional excitation in each
cavity requires energy 
$$ {E_2 - 2 E_1} = 2 \sqrt{ {\lambda}^{2} ~ +~ \frac{{\Delta}^2 }{4} }
- \sqrt{2 {\lambda}^{2} ~ +~ \frac{{\Delta}^2 }{4} } - \frac{\Delta}{2}. $$ 
Which one may consider as an effective on-site repulsion because it measures
the difference between the energy of two and single excitation (polariton in
each cavity). 
This effective repulsion
decreases as we increase the detune factor. Therefore we can conclude that
for $\Delta =0 $, double occupation never occurs, indicating a Mott insulating
behaviour. When $ \Delta $ is much larger than the coupling $\lambda $ then
the occupation number larger than one occurs as to be expected for a 
photonic superfluid
regime. In our quantum field theoretical calculations, we have also predicted
that the large detuning drives the system from gapped Mott insulating phase 
to the gapless superfluid phase.\\
\section{Conclusions}
At first we have done the quantum field theoretical analysis of an effective
spin model in coupled optical cavity arrays. We have predicted two quantum
phases, Mott insulator and photonic superfludity. 
Anisotropy in the exchange interaction
has also created a gapped phase. An infinite order KTB transition has
been replaced by the second order XY transition. The rigorous
quantum field theoretical derivation of this manuscript is absent in all 
previous studies and also we provide physical explanation of the transition
process based on Jaynes-Cummings Hamiltonian.

\centerline{\bf Acknowledgments}
\vskip .2 true cm
The author would like to acknowledge The Center for Condensed Matter 
Theory of the 
Physics Department
of IISc for providing working space and also Dr. R. C. Sarasij for reading
the manuscript very critically.
 
\begin{figure}
\includegraphics[scale=1.5,angle=0]{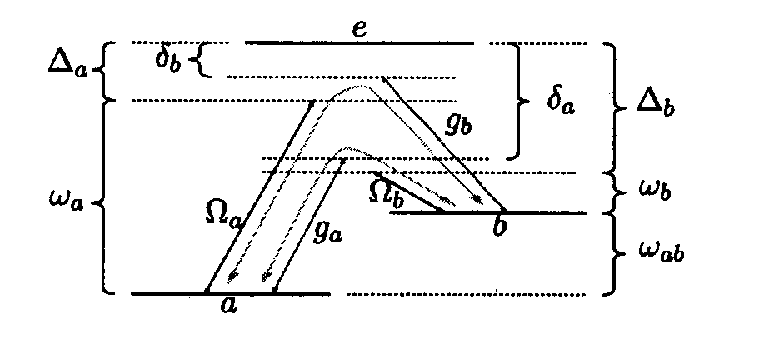}
\caption{Level structure, driving laserrs, and relevant couplings to
the cavity mode to generate effective ${{\sigma}^{x}}{{\sigma}^{x}}$
and ${{\sigma}^{y}}{{\sigma}^{y}}$ couplings for one atom. The cavity
mode couples with strength ${g_a}$ and ${g_b}$ to transition 
$|a> \leftrightarrow |e>$ and $|b> \leftrightarrow |e>$ respectively.
One laser frequency ${\omega}_a $ couples to the transition 
$|a> \leftrightarrow |e>$ with Rabi frequency 
$ {\Omega}_a $ and another with frequency ${\omega}_b $ to
$|b> \leftrightarrow |e>$ with $ {\Omega}_b $. The dominant two-photon
processes are indicated in faint arrows. Reprinted with permission from
American Physical Society. Analytical expressions and physical meaning of
different symbols have given in Ref.\cite{hart2}. }  
\label{Fig. 1 }
\end{figure}
\begin{figure}
\includegraphics[scale=1.5,angle=0]{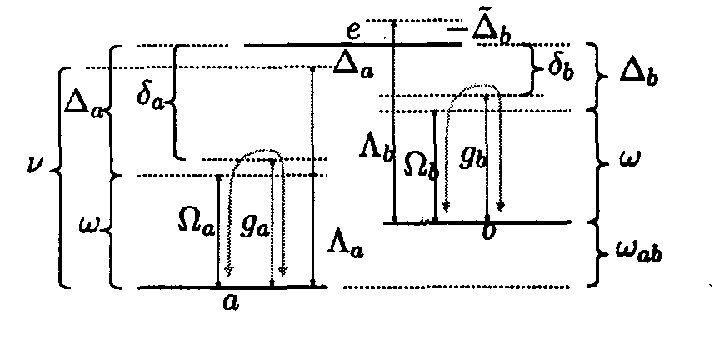}
\caption{Level structure, driving laserrs, and relevant couplings to
the cavity mode to generate effective ${{\sigma}^{z}}{{\sigma}^{z}}$
couplings for one atom. The cavity
mode couples with strength ${g_a}$ and ${g_b}$ to transition
$|a> \leftrightarrow |e>$ and $|b> \leftrightarrow |e>$ respectively.
Two lasers with frequencies $\omega $ and $\nu$ couple with Rabi frequencies 
$ {\Omega}_a $, respectively, ${\Lambda}_a $ to transition
$|a> \leftrightarrow |e>$ and
$ {\Omega}_b $, respectively, ${\Lambda}_b $ to transition
$|b> \leftrightarrow |e>$
. The dominant two-photon
processes are indicated in faint arrows.
Reprinted with permission from
American Physical Society. Analytical expressions and physical meaning of
different symbols have given in Ref.\cite{hart2}.
}
\label{Fig. 2 }
\end{figure}


\begin{references}
\bib{rao} C. N. R. Rao and T. V. Ramakrishnan in {\it  
Superconductivity Today} (Universities Press, Hyderabad, 1999).

\bib{lik1} K. K. Likharev in {\it Dynamics of Josephson junction and circuits}
(Gordon and Breach 1988).

\bib{dens} J. Hecker. Denschlog $et~al.$, J. Phys. B: At. Mol. Opt Phys,
{\bf 35}, 3095 (2002). 

\bib{zoller} D. Jaksch and P. Zoller, Annals of Physics {\bf 315}, 52 (2005).
\bib{green} A. D. Greentree $et~al.$, Nature Phys. {\bf 466}, 856 (2006).

\bib{hart1}  Michael J. Hartmann, Fernando G. S. L Brando and Martin B. Plenio,
Nature Phys. {\bf 462}, 849 (2006);
Michael J. Hartmann, Fernando G. S. L Brando and Martin B. Plenio, Laser and 
Photonics Rev. {\bf 2}, 527 (2008). 

\bib{hart2} Michael J. Hartmann, Fernando G. S. L Brando and Martin B. Plenio,
Phys. Rev. Lett {\bf 99}, 160501 (2007).

\bib{ji} A-C.Ji, X. C. Xie, and W. M. Liu, Phys. Rev.Lett. {\bf 99},
183602 (2007).

\bib{byrn} T. Byrnes, N. Y. Kim, K. Kusudo, and Y. Yamamoto, 
Phys. Rev. B {\bf 78}, 075320 (2008).

\bib{caru} I. Carusotto $et~al.$, arXiv:0812.4195 (2008).

\bib{bhas} M. J. Bhaseen, M. Hohenadler, A. O. Silver, and B. D. Simons,
Phys. Rev. Lett. {\bf 102}

\bib{toma} A. Tomadin $et~al.$ arXiv:0904.4437 (2009).

\bib{zhao} J. Zhao, A. W. Sandvik, and K. Ueda, arXiv:0806.3603 (2008).

\bib{pipp} P. Pippan, H. G. Evertz, and M. Hohenadler, arXiv: 0904.1350 (2009).

\bib{rossi} D. Rossini, and R. Fazio, Phys. Rev. Lett. {\bf 99}, 186401 (2007).

\bib{horn}  M. Aichhorn $et~al.$, Phys. Rev. Lett. {\bf 100}, 216401 (2008).

\bib{blat} S. Schmidt, and G. Blatter, arXiv:0905.3344 .

\bib{bose}  D. G. Angelakis, M. F. Santos and S. Bose, Phys. Rev. A {\bf 76},
R031805 (2007).

\bib{gia} T. Giamarchi in {\it Quantum Physics in One Dimension} (Clarendon
Press, Oxford 2004).

\bib{zamo} S. Lukyanov and A. Zamolodchikov, Nucl. Phys. B {\bf 493}, 571 (1997).

\bib{phase} In quantum phase transitions, in addition to the standard critical
exponent it is useful to define an additional exponent z, called the 
dynamical exponent, which tells us how a characteristic length in the time 
direction is related to a length in the spatial direction 
${\xi}_{\tau} \sim {{\xi}_{x}}^{z}$. For quantum problem time plays a
special role and this special direction has no reasons to have the same
exponent as the spatial one.  
Deep inside the Mott insulating phase, particle and hole
excitations are gapped. The system is almost alike to the atomic limit in
the deep Mott insulating phase. When one approaches the phase boundary from the
deep Mott insulating phase then the dispersion relation is quadratic 
${\omega} \sim {k^2} $ ($z= 2 $). But the situation is different
at the end of CDW phase and the starting point of XY staggered order phase.
At around the multical critical point $z$ is $1$. Please see the 
Refs. \cite{subir,sondhi} and 
for a detailed understanding of this subject.

\bib{subir} Subir Sachdev in "Quantum Phase Transition" (Cambridge University
Press Cambridge, 1998).

\bib{sondhi} S. L. Sondhi, $et~al.$ Rev. Mod. Phys. {\bf 69}, 315 (1997).

\bib{cardy} J. Cardy in {\it Scaling and Renormalization in Statistical Physics}
(Cambridge University Press, Cambridge 1996); I. Affleck in 
{\it Fields, Strings and and Critical Phenomena}, ed E. Brezin and J. Zinn-Justin
(North-Holland, Amstardam 1989).

\bib{jay} Jaynes, E. T. and F. W. Cummings, Proc IEEE, {\bf 51}, 89 (1963).

\bib{horo} S. Horoche and J. M. Raimond in {\it Exploring the Quantum Atoms,
Cavities, and Photons}, (Oxford University Press, 2006).

 
\end{references}
\end{document}